\documentclass[doublespacing]{elsart}

\usepackage{graphics}
\usepackage{graphicx}
\usepackage{epsfig}

\usepackage{subfigure}

\usepackage{amssymb}
\usepackage{amsmath}
\usepackage{amsfonts}
\usepackage{stmaryrd}
\usepackage{txfonts}
\usepackage{mathrsfs}
\usepackage{latexsym, bm}
\usepackage{CJK}
\newtheorem{theorem}{Theorem}

\newtheorem{property}{Property}
\newtheorem{definition}{Definition}
\newtheorem{lemma}{Lemma}
\newtheorem{corollary}{Corollary}

\newtheorem{remark}{Remark}

\newtheorem{example}{Example}
\newtheorem{assumption}{Assumption}

\DeclareMathOperator{\diag}{diag}
\DeclareMathOperator{\adj}{adj}
\journal{}
\begin{document}

\begin{frontmatter}


\title{Optimal topology of multi-agent systems with two leaders: a zero-sum game perspective \thanksref{label1}}
\thanks[label1]{This work was supported by 973 Program (Grant No. 2012CB821203), NSFC (Grant
Nos. 61020106005, 61375120 and 61304160) and the Fundamental Research Funds for
the Central Universities (Grant No. JB140406). Bin Wu acknowledges the sponsorship from Max-Planck society.}
\author[Xidian]{Jingying Ma}, \ead{majy1980@126.com}
\author[Xidian]{Yuanshi Zheng}, \ead{zhengyuanshi2005@163.com}
\author[Germany]{Bin Wu},\ead{bin.wu@evolbio.mpg.de}
\author[Peking]{Long Wang\corauthref{cor1}}\ead{longwang@pku.edu.cn}
\corauth[cor1]{Corresponding author : Long Wang }
\address[Xidian]{Center for Complex Systems, School of Mechano-electronic Engineering,\\
Xidian University, Xi'an 710071, P.~R.~China}
\address[Germany]{Department of Evolutionary Theory, Max-Planck-Institute for Evolutionary Biology,\\
August-Thienemann-Str. 2, 24306 Pl\"on, Germany}
\address[Peking]{Center for Systems and Control, College of Engineering, \\
Peking University, Beijing 100871, P.~R.~China}
\begin{abstract}
It is typical to assume that there is no conflict of interest among leaders. Under such assumption, it is known that, for a multi-agent system with two leaders, if the followers' interaction subgraph is undirected and connected, then followers will converge to a convex combination of two leaders' states with linear consensus protocol. In this paper, we introduce the conflict between leaders: by choosing $k$ followers to connect with, every leader attempts all followers converge to himself closer than that of the other. By using graph theory and matrix theory, we formulate this conflict as a standard two-player zero-sum game and give some properties about it. It is noteworthy that the interaction graph here is generated from the conflict between leaders. Interestingly, we find that to find the optimal topology of the system is equivalent to solve a Nash equilibrium. Especially for the case of choosing one connected follower, the necessary and sufficient condition for an interaction graph to be the optimal one is given. Moreover, if followers' interaction graph is a circulant graph or a graph with a center node, then the system's optimal topology is obtained. Simulation examples are provided to validate the effectiveness of the theoretical results.
\end{abstract}
\begin{keyword}
multi-agent systems; zero-sum game; containment control; optimal topology
\end{keyword}
\end{frontmatter}
\section{\bf Introduction}\label{s1}
A multi-agent system is composed of multiple autonomous agents which can exchange information and mutually interact. In recent years, distributed control of multi-agent systems(MASs) have attracted intensive attention in the scientific community. This is due to its diverse applications in many areas, such as formation control in unmanned aerial vehicles\cite{Feng Xiao, game2}, flocking in biology \cite{Flocking}, coverage control of sensor networks \cite{sensor2}, attitude alignment of satellite clusters \cite{Ren W3}, and so on.

\subsection{Related works}
Consensus seeking is a basic problem of MASs which aims to design appropriate distributed protocols or algorithms such that a group of agents can converge to the same state. It was originally studied by Vicsek \emph{et al.} \cite{Vicsek} and was theoretically explained by graph theory in \cite{Jabdabaie}. Olfati-Saber and Murray developed a systematical framework of consensus problem in networks of dynamic agents with switching topology and time-delays in \cite{Saber}. Some relaxed conditions were obtained for first-order MASs in \cite{Ren W2}, where consensus is solved if there exists a spanning tree. Following \cite{Jabdabaie,Saber,Ren W2}, there have been extensive studies and results under various circumstances, to name but a few, second-order consensus \cite{Xie G}, consensus of heterogeneous MASs \cite{Zheng Y1,Zheng Y3} and finite-time consensus \cite{WangLong 2, Zheng Y4}, etc.

As a special role in MASs, leaders are ubiquitous in nature, for instance, the navigation aircraft in a fight formation of UAVs and the leading whale in a whale population. This fact attracts researchers' great attention and leads to some research hotspots such that leader-following consensus \cite{Yiguang Hong,Daizhan Cheng}, containment control problem \cite{Huiyang Liu,Zheng Y5} and controllability analysis \cite{control1,control2,control3}. For a multi-agent system with a single leader, followers will converge to the state of leader, which is called the tracking control or leader-following consensus problem. Sufficient conditions for solving leader-following consensus were brought up for MASs \cite{Yiguang Hong,Daizhan Cheng}. Based on linear quadratic regular theory, Ma \emph{et al.} proved that the optimal topology of leader-following consensus is a star graph \cite{Majingying}. As an extension of leader-following consensus, containment control problem of MASs with multiple leaders means that the states of the follower will converge to the convex hull spanned by the leaders. In \cite{containment1}, the authors assumed the leaders located in vertices of a convex polytope, and presented a hybrid Stop-Go strategy for first-order leaders with the fixed undirected topology. Notarstefano \emph{et al.} investigated containment control of first-order MASs with switching topologies \cite{containment2}. Liu \emph{et al.} obtained the necessary and sufficient conditions for solving containment control of multi-agent systems with multiple stationary and dynamic leaders under directed networks in \cite{Huiyang Liu}.  Some research topics of containment control under different systems have also been addressed, such as containment control for second-order MASs \cite{containment4}, for heterogeneous MASs \cite{Zheng Y5} and for MASs with measurement noises \cite{Zheng Y}.

In a multi-agent system, each agent is an individual who exchanges information with their neighbors and then makes decision independently. If we further define the utility of agents and assume that individuals adjust its behaviors by promoting utility, game theory can be introduced to distributed multi-agent coordination. By reviewing consensus seeking as a non-cooperative differential game, Bauso \emph{et al.}\cite{Bauso} proposed a game theoretic interpretation of consensus problems as mechanism design problems. By imposing individual objectives, the author proved that
such objectives can be designed so that rational agents have a unique optimal protocol, and asymptotically reach consensus on a desired group
decision value. In \cite{game2},  the author investigated formation control via a linear-quadratic (LQ) Nash differential game and gave a RHC-based approach.  
While as in \cite{Yilang3}, cooperative game theory is utilized to ensure team cooperation by considering a combination of individual cost as the team cost and the Nash-bargaining solution is obtained. For leader-following MASs, the notion of graphical game was formulated in \cite{Lewis4}. The author brought together cooperative control, reinforcement
learning, and game theory to solve multi-player differential games on communication graph topologies and proposed a cooperative policy iteration algorithm for graphical games that converges to the best response. Gharesifard and Cort\'{e}s introduced the distributed convergence to Nash equilibrium for two networks engaged in a strategic scenario in \cite{game3}.

\subsection{Our results}
Different from the aforementioned literatures, we propose a multi-agent system with two leaders and formulate a type of game. For a multi-agent system with two leaders, if the followers' interaction subgraph $G_F$ is undirected and connected, then each follower will converge to a convex combination of two leaders' states \cite{Huiyang Liu}. Based on this result, we assume that every leader can independently select $k ~(\geq1)$ followers to connect with him. Then, we define the average distance to the followers as the payoff function of each leader. There is a conflict of interest between two leaders -- what one gains incurs a loss to the other. Therefore, we can describe this process as a noncooperative game in which each leader independently chooses an optimal strategy (i.e., the connected followers) to minimize his payoff function. Noticing that two leaders' decisions will determine the interaction topologies of the system, a Nash equilibrium point corresponds to an optimal topology of the system. The main contributions of this paper are threefold. Firstly, we formulate a new type of game for multi-agent system. Secondly, by utilizing containment control and matrix theory, we reformulate the game as a zero-sum game denoted by $\mathcal{G}^k(G_F)$ and develop some properties based on game theory.
Finally, for the case $k=1$, the necessary and sufficient condition for an interaction graph to be the optimal topology is given. Moreover, if $G_F$ is a circulant graph or a graph with a center node, then the optimal topology is obtained. It should be mentioned that these results offer some theoretical explanations to some commercial and political phenomena. Consider two companies selling similar product or two candidates promoting an election, both of two opponents aim to propagate their opinion in social networks by choosing some members as their supporters from it. In the scenario where the influence power of every member is equal, everyone is the optimal strategy for two opponents. If a social network exists a 'center' member who can influence all the others, then both of two opponents will select this member to maximize their own influence.

This paper is organized as follows. In Section \ref{s2}, we introduce the graph theory and two-person zero-sum game and propose our problem. In Section \ref{s3}, we give our main results. And in Section \ref{s4}, numerical simulations are given to illustrate the effectiveness of the theoretical results. Some conclusions are drawn in Section 5.

\textbf{Notation:} Throughout this paper, the following notations will be used: let $\mathbb{R}$ be the set of real numbers. $\mathbb{R}^{n\times m}$ is the set of $n\times m$ real matrices.
Denote by $\mathbf{1}_n$ (or $\mathbf{0}_n$) the column vector with all entries equal to one (or all zeros). $I_n$ denotes an $n-$dimensional identity matrix. For a column vector $\mathbf{b}=[b_1,b_2,\dots,b_n]^T$, $\diag\{\mathbf{b}\}$ is a diagonal matrix with $b_i$, $i=1,...,n,$ on its diagonal and $\parallel \mathbf{b} \parallel_1=\sum_{i=1}^n|b_i|$ is 1-norm of $\mathbf{b}$. For a matrix $A \in \mathbb{R}^{n \times n}$, $\adj A$ and $\det A$ are the adjugate and the determinant of $A$, respectively. Denote $A(i,j)$ as the matrix obtained by deleting row $i$ and column $j$ from $A$ and $A_{[i_1,i_2,...,i_k]}$ be the $k\times k$ principal submatrix of $A$ by keeping rows and columns $i_1,i_2,...,i_k$. $\mathcal{I}_n=\{1,\ \dots,\ n\}$ is an index set. $|S|$ is the cardinality of a set $S$. For two sets $S_1$ and $S_2$, denote $S_1\times S_2$ as the Cartesian product and $S_1\setminus S_2=S_1-S_2$. Let $\mathbf{e}_i=[e_1,e_2,\dots,e_n]^T$ is the unit column vector with $e_i=1$ and $e_j=0$, $j\neq i$. The notation $A \Leftrightarrow B$ means that $A$ holds on if and only if $B$ holds on.
\section{\bf Preliminary}\label{s2}
\subsection{Graph Theory }
In this subsection, we present some basic notions of algebraic graph which will be used in this paper.

Let $G=\{V,E\}$ be an undirected graph consisting of a vertex set $V=\{1,2,...,n\}$ and an edge set $E=\{(i,j)\in V\times V\}$. If $(i,j)\in E$, then we say that $i$ and $j$ are adjacent or  that $j$ is a neighbor of $i$. For a subset of $V$, denoted by $W$, a graph $G_W=\{W,E(W)\}$ is an induced subgraph of $G$ if two vertices of $W$ are adjacent in $G_W$ if and only if they are adjacent in $G$. The adjacency matrix $A$ of $G$ is a symmetric $01$-matrix such that for all $i\in\mathcal{I}_n$, $a_{ii}=0$ and for all $i\neq j$, $(i,j)\in E \Leftrightarrow a_{ij}=a_{ji}=1$, while $a_{ij}=0$ otherwise. The neighbor set of the vertex $i$ is $\mathcal{N}_i=\{j: (i,j)\in E\}$. The degree matrix $D \in \mathbb{R}^{n\times n}$ is a diagonal matrix with $d_{i}=\sum_{j\in \mathcal{N}_i }a_{ij}$ and the Laplacian matrix $L=D-A$. A vertex $i$ is a center node if it connects all the other vertexes, i.e., $\mathcal{N}_{i}=V\setminus\{i\}$. A graph is called a star graph if there exists a center vertex denoted as $i^c$, and for all $j \in V\setminus\{i^c\}$, $\mathcal{N}_j=\{i^c\}$. A complete graph means that every pair of vertices are adjacent. A graph is circulant if the adjacency matrix is a circulant matrix. It is obviously that a complete graph is circulant.
\begin{lemma}\label{e-1}\cite{AGT}
For a graph $G$,
$\det L=0$ and $\adj L=\tau(G)\mathbf{1}_n\mathbf{1}_n^T$, where $\tau(G)$ is the number of spanning trees in the graph $G$.
\end{lemma}
\begin{lemma} \label{lemma L}
If a graph $G$ is connected, then every principal submatrix of $L$ is positive definite and moreover, the inverse matrix of it is a nonnegative matrix.
\end{lemma}

{\it Proof.} Denote $W=\{i_1,i_2,...,i_k\}$ be a subset of $V$, and $G_{W}=\{W,E(W)\}$ be the induced subgraph of $G$ where $1\leq i_1<i_2,...<i_k\leq n$. Then it follows that the adjacent matrix of $G_W$, denoted by $A_W$, is the principal submatrix of $A$ by retaining rows and columns $i_1,i_2,...,i_k$. Let $L_W$ be the Laplacian matrix of $G_W$, $\theta_{i_m}=-\sum_{j\in W} l_{i_mj}$, $m=1,2,...,k$. Thus we have $L_{[i_1,i_2,...,i_k]}=L_W+\diag\{\theta_{i_1},...,\theta_{i_k}\}$. Therefore, from Lemma 4 in \cite{Daizhan Cheng}, we obtain that $L_{[i_1,i_2,...,i_k]}$ is positive definite. Moreover, we have
\[L_{[i_1,i_2,...,i_k]}=\diag\{l_{i_1i_1},...,l_{i_ki_k}\}-A_W
=\eta \left(I_n- (\Delta +\frac{A_W}{\eta} )\right)\]
where $\eta=\max_{1\leq m\leq k} l_{i_mi_m}$ and $\Delta=\frac{1}{\eta }\diag\{\eta-l_{i_1i_1},...,\eta -l_{i_ki_k}\}$. Hence, it follows that
\[L_{[i_1,i_2,...,i_k]}^{-1}=
\frac{1}{\eta }\sum_{k=0}^{\infty}\left(\Delta +\frac{A_W}{\eta} \right)^k.\]
Because $\Delta$ and $\frac{A_W}{\eta}$ are nonnegative matrices, $L_{[i_1,i_2,...,i_k]}^{-1}$ is a nonnegative matrix. $\blacksquare$

\subsection{Two-person zero-sum games}
In this subsection, we present the notions of a class of two-player zero-sum games where each player has a finite number of strategies to choose from. For more details, please refer to \cite{Basar T}.

Consider a zero-sum game of two players, to be referred to as player $P_1$ and player $P_2$, in which each player has finite alternatives. Denote the set of strategies of $P_1$ and $P_2$ as $S_1=\{s_1,s_2,...,s_m\}$ and $S_2=\{\hat{s}_1,\hat{s}_2,...,\hat{s}_n\}$, respectively. A pair of strategies $(s_i,\hat{s}_j) \in S_1\times S_2$ means that $P_1$ chooses the strategy $s_i$
and $P_2$ chooses the strategy $\hat{s}_j$. For strategies pair $(s_i,\hat{s}_j)$, the payoff of $P_1$ is $-a_{ij}$ while that of $P_2$ is $a_{ij}$. Then, $A=\{a_{ij}\}_{m\times n}$ is called the outcome of the game and this type of two-person zero- sum game is called a matrix game $A$. In a matrix game $A$, $P_1$ wants to minimize the outcome of the game, while $P_2$ seeks to maximize it, by independent decisions. Under such an incentive, $P_1$ is forced to pick a strategy $s_{i^*}$ satisfied
\[\overline{V}(A)\triangleq \max _ja_{i^*j}=\min_i \max _ja_{ij}.\]
The strategy $s_{i^*}$  is called a security strategy for $P_1$. Similarly, $P_2$ will choose a security strategy $\hat{s}_{j^*}$ determined by
\[\underline{V}(A)\triangleq \min _ia_{ij^*}=\max_j \min _ia_{ij}.\]
Denote $S_1^*$ and $S_2^*$ be the set of the security strategies of $P_1$ and $P_2$, respectively.
\begin{lemma}\cite{Basar T}
In every matrix game $A =\{a_{ij}\}$,
\begin{enumerate}
  \item $\overline{V}(A)$ and $\underline{V}(A)$ are unique,
  \item there exists at least one security strategy for each player,
  \item  $\underline{V}(A)\leq \overline{V}(A)$.
\end{enumerate}
\end{lemma}
\begin{definition}\cite{Basar T}
For a given $(m\times n)$ matrix game $A =\{a_{ij}\}$, if a strategies pair $({s_{i^*}, \hat{s}_{j^*}})$ satisfied $a_{i^*j}\leq a_{i^*j^*}\leq a_{ij^*}$
for all $i \in \mathcal{I}_m$ and $j \in \mathcal{I}_n$, then it is said that the matrix game has a Nash equilibrium point in pure strategies. The corresponding outcome $a_{i^*j^*}$ of the game is called the Nash equilibrium outcome denoted by V(A).
\end{definition}
\begin{lemma}\cite{Basar T}\label{lemma2}
Let $A =\{a_{ij}\}$ denote an $(m\times n)$ matrix game with $\underline{V}(A)=\overline{V}(A)$. Then,
\begin{enumerate}
  \item $A$ has a (pure) Nash equilibrium point,
  \item the strategies pair $(s_i,\hat{s}_j)$ is a Nash equilibrium point for $A$ if and only
if $s_{i}\in S_1^*$ and $\hat{s}_{j} \in S_2^*$.
  \item V(A) is uniquely given by $\underline{V}(A)=\overline{V}(A)$.
\end{enumerate}
\end{lemma}
\subsection{Problem statement}
Consider a multi-agent system consisting of $n$ followers and two leaders. The set of the followers is denoted as $V=\{1, \dots , n\}$, and the two leaders are denoted as $l_0$ and $l_1$. The interaction of the followers is described by an undirected graph $G_F=(V,E)$. The following assumption is given throughout this paper.
\begin{assumption}
 (Connectivity) $G_F$ is connected.
\end{assumption}
The leaders $l_0$ and $l_1$ keep static states denoted by $y_0,~y_1 \in R$, respectively. Without loss of generality, we assume $y_0<y_1$. The state of follower $i \in V$ is denoted as $x_i(t)\in R$. The dynamics of $x_i(t)$ is given by
\begin{equation}\label{e1}
\dot x_i=\sum_{j \in \mathcal{N}_i}(x_j-x_i)+b_i(y_0-x_i)+d_i(y_1-x_i)
\end{equation}
where $\mathcal{N}_i$ represents the neighbor set of $i$ in $G_F$ and
\[b_i=\left\{\begin{aligned}
               &1, ~~if~~i~~is~~connected~~to~~l_0 \\
              &0,~~otherwise
             \end{aligned}
\right.~~\text{and}~~
d_i=\left\{\begin{aligned}
               &1, ~~if~~i~~is~~connected~~to~~l_1 \\
              &0,~~otherwise
             \end{aligned}
\right.\]
denote whether the follower $i$ is connected to the leader $l_0$ and $l_1$ or not, respectively. Denote $\mathbf{b}=[b_1,\dots,b_n]^T$, $\textbf{d}=[d_1, \dots,d_n]^T$ and $X_f(t)=[x_1(t), \dots, x_n(t)]^T$, then it follows that
\begin{equation}\label{e2}
\dot X_f(t)=-(L+\diag\{\textbf{b}+\textbf{d}\})X_f(t)+\textbf{b}y_0+\textbf{d}y_1\\
\end{equation}
where $L$ is the Laplacian matrix of $G_F$. Form \cite{Huiyang Liu}, we have
\begin{lemma}\label{lemma3}
If each leader has connected one agent at least in $G_F$ (i.e., $\mathbf{b} \neq \mathbf{0}_n$ and $\mathbf{d} \neq \mathbf{0}_n$), then $X_f(t)$ will converge to
\[\lim_{t\rightarrow \infty}X_f(t)=\alpha y_0+\beta y_1\]
where
\begin{equation}\label{e3}
\begin{aligned}
&\alpha=[\alpha_1,\alpha_2,...,\alpha_n]^T=(L+\diag\{\mathbf{b}+\mathbf{d}\})^{-1}\mathbf{b},\\
&\beta=[\beta_1,\beta_2,...,\beta_n]^T=(L+\diag\{\mathbf{b}+\mathbf{d}\})^{-1}\mathbf{d},\\
\end{aligned}
\end{equation}
$\alpha_i +\beta_i=1$, $\alpha_i>0$ and $\beta_i>0$ for all $i \in V$.
\end{lemma}
In this paper, we consider the following game regarding the multi-agent system (\ref{e2})(see Fig. \ref{fig1}):
\begin{figure}[htbp]
  \centering
  \includegraphics[scale=0.6]{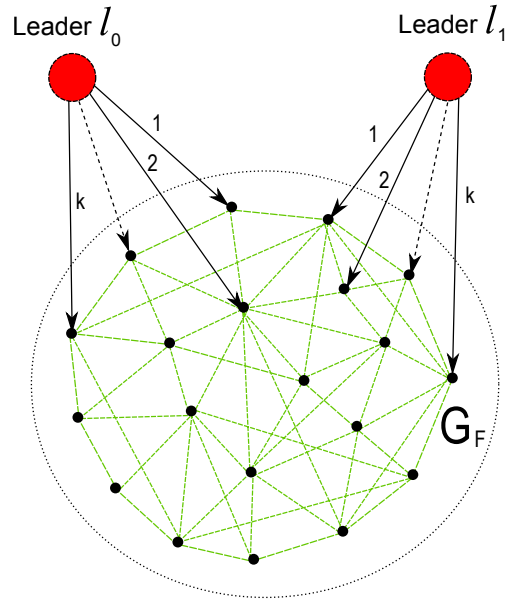}\\
  \caption{An interaction graph $\tilde{G}(s_i, s_j)$ determined by the strategies pair $(s_i, s_j)$ }\label{fig1}
\end{figure}
\begin{description}
  \item[Players] Let $l_0$ and $l_1$ be two players.
  \item[Strategies] Each player can select $k$ ($1 \leq k \leq n$) followers from $V$ to connected with, i.e., the set of strategies of each player is $S=\{s_j=(a_1,a_2,...,a_n)^T\mid a_i\in\{0,1\},\sum_{i=1}^na_i=k\}$. Obviously, $S$ is a finite set and $|S|=C_n^k\triangleq N$. Then, let $S=\{s_1,s_2,...,s_N\}$.
  \item[Payoff] The goal of each player is to steer all followers to move forward to himself as closely as possible. As a result, the payoff of each leader can be described as the average distance between the followers and himself. Define
      \[\begin{aligned}
      &U_0(s_i,s_j)=\frac{1}{n}\sum_{m=1}^{n}\mid \lim_{t\rightarrow \infty}x_m(t)-y_0\mid,\\
      &U_1(s_i,s_j)=\frac{1}{n}\sum_{m=1}^{n}\mid \lim_{t\rightarrow \infty}x_m(t)-y_1\mid,
      \end{aligned}\]
     as the payoff function of $l_0$ and $l_1$, respectively. Then every player wants to chooses a strategy from $S$ to minimize his payoff.
\end{description}
According to Lemma \ref{lemma3}, we find
\begin{equation}\label{e5}
U_0(s_i,s_j)=\frac{1}{n}\sum_{k=1}^{n}\beta_k(y_1-y_0), ~~U_1(s_i,s_j)=\frac{1}{n}\sum_{k=1}^{n}\alpha_k(y_1-y_0)
\end{equation}
and
\begin{equation}\label{e4}
U_0(s_i,s_j)+U_1(s_i,s_j)=y_1-y_0
\end{equation}
for all $(s_i,s_j)\in S \times S$. Since $y_1-y_0$ is a constant, there is a conflict of interest between two leaders -- what one leader gains incurs a loss to the other leader. Therefore, this game is noncooperative.
\begin{remark}
It should be mentioned that this game implies that each leader independently selects his own edges to minimize his payoff. Then, for system (\ref{e2}), the two leaders' choices $(s_i, s_j)$ determine its interaction graph which consists of follower interaction graph $G_F$, two leader vertexes $l_0$ and $l_1$ and $2k$ directed edges from the leaders to the followers (see Fig. \ref{fig1}). Denoted this graph as $\tilde{G}(s_i, s_j)$. If a strategies pair $(s_{i^*}, s_{j^*})$ minimizes both two leader's payoff functions, then the corresponding interaction graph $\tilde{G}(s_{i^*}, s_{j^*})$ is called the optimal topology of the system (\ref{e2}).
\end{remark}
\begin{remark}
This game is common in real world, especially in human society. Consider an example from commercial world. Two companies
sell similar products in a market. A consumer will select the company whose product is 'closer' to him in taste or fashion. And each company will select some spokesmen to promote his product. The spokesman usually will exert great influence on the consumer in his decision making. Suppose Cristiano Ronaldo is the spokesman of Nike, then it is more likely that his fans will choose Nike because of him. The product is thus closer to him due to the successful marketing strategy. Election issue can also be modeled by this game. Consider two candidates run for election. A voter will vote the candidate closer to him in opinion. Assume the dynamics of voter's opinion is determined by his neighbors in social networks. Then, aiming to win the election, each candidate seeks to find some loyal supporters to broadcast his politics opinion.
\end{remark}

\section{\bf Main results}\label{s3}
In this section, we will first reformulate the above game and give some properties about it. Then the case of $k=1$ will be investigated further, and the necessary and sufficient condition for an interaction graph to be an optimal topology will be gotten. Moreover, the optimal topologies of some special cases will be solved.
\subsection{Probelm reformulation }

It follows from (\ref{e3}) and (\ref{e4}) that
\[
\begin{aligned}
&\min_{s_i\in S}U_0(s_i,s_j)
=(y_1-y_0)\min_{s_i\in S}\sum_{k=1}^{n}\frac{\beta_k}{n},\\
&\min_{s_j\in S}U_1(s_i,s_j)
=(y_1-y_0)\left(1-\max_{s_j\in S}\sum_{k=1}^{n}\frac{\beta_k}{n}\right)
\end{aligned}
\]
and
\begin{equation}\label{e6}
\sum_{k=1}^{n}\frac{\beta_k}{n}
=\frac{1}{n}\mathbf{1}_n^T(L+\diag\{s_i+s_j\})^{-1}s_j.
\end{equation}
Therefore, $l_0$ attempts to minimize $\sum_{k=1}^{n}\frac{\beta_k}{n}$, while $l_1$ intends to maximize it, by independent decisions. Then we can formulate this game as a two-person zero-sum game. From (\ref{e6}), it is noticed that $\sum_{k=1}^{n}\frac{\beta_k}{n}$ is independent of the initial states of the followers and the leaders and is only determined by the structure of $G_F$ and the strategies pair $(s_i, s_j)$. Then we can redefine the game as following:
\begin{definition}
For a connected graph $G_F$, there are two players, $l_0$ and $l_1$, and an $N\times N$ matrix $U=\{u_{ij}\}_{N\times N}=\{\frac{1}{n}\mathbf{1}_n^T(L+\diag\{s_i+s_j\})^{-1}s_j\}$, $s_i,s_j\in S$.  The player $l_0$ wants to pick a strategy $s_i \in S $ satisfied
\[\min_{s_i\in S}\max_{s_j\in S}u_{ij},\]
and the player $l_1$ seeks to choose a strategy $s_j \in S $ such that
\[\max_{s_j\in S}\min_{s_i\in S}u_{ij}.\]
Then this matrix game is denoted as $\mathcal{G}^k(G_F)$ and $U$ is the outcome matrix.
\end{definition}
\begin{definition}\label{optimal}
For the game $\mathcal{G}^k(G_F)$, if a strategies pair $(s_{i^*}, s_{j^*})$ is a Nash equilibrium point, i.e., $u_{i^*j}\leq u_{i^*j^*}\leq u_{ij^*}$, then the interaction graph $\tilde{G}(s_{i^*}, s_{j^*})$
is optimal for $l_0$ as well as for $l_1$. We say that $\tilde{G}(s_{i^*}, s_{j^*})$ is the optimal topology of the system (\ref{e2}).
\end{definition}
In the subsequent development, some properties of the game $\mathcal{G}^k(G_F)$ will be investigated.
\begin{property}\label{Pro0}
For the strategies pair $(s_i,s_j)$, one has $U_0(s_i,s_j)=u_{ij}(y_1-y_0)$ and $U_1(s_i,s_j)=(1-u_{ij})(y_1-y_0)$. Moreover, $u_{ij}<(=~ or ~>)~\frac{1}{2}$ if and only if
$U_0(s_i,s_j)<(=~ or ~>)~U_1(s_i,s_j)$.
\end{property}
{\it Proof.} The proof is straightforward from the definition of $U$ and (\ref{e5}). $\blacksquare$

\begin{property}\label{Pro1}
 $U+U^T=\mathbf{1}_N^T\mathbf{1}_N$, i.e., $u_{ij}+u_{ji}=1$ for all $i,j \in \mathcal{I}_N$. Moreover, $\overline{V}(U)\geq \frac{1}{2}$ and $\underline{V}(U)\leq  \frac{1}{2}$.
\end{property}
{\it Proof.}
According to Lemma \ref{lemma3}, one has $(L+\diag\{s_i+s_j\})^{-1}(s_i+s_j)=\mathbf{1}_n$. Then it follows that
\[u_{ij}+u_{ji}=\frac{1}{n}\mathbf{1}_n^T(L+\diag\{s_i+s_j\})^{-1}(s_i+s_j)
=\frac{1}{n}\mathbf{1}_n^T\mathbf{1}_n=1.\]
Then we get $u_{ii}=\frac{1}{2}$ for all $i \in \{1,2,...,N\}$. Hence, $\overline{V}(U)=\min_i \max _ja_{ij}\geq \frac{1}{2}$ and $\underline{V}(U)=\max _j\min_ia_{ij}\leq \frac{1}{2}$. $\blacksquare$

\begin{remark}
If there is a strategy $s_{i^*}$ such that $u_{i^*j}\leq \frac{1}{2}$ for all $j\in V$, then $\overline{V}(U)=\underline{V}(U)=\frac{1}{2}$.
\end{remark}

\begin{property}\label{pro3}
Let $S_0^*$ and $S_1^*$ be the security strategies set of $l_0$ and $l_1$, respectively. Then $S_0^*=S_1^*=S^*$. Moreover, if $\overline{V}(U)=\frac{1}{2}$, then $(s_{i},s_{j})$ is a Nash equilibrium point if and only if $(s_{i},s_{j})\in S^*\times S^*$, and the Nash equilibrium outcome is $\frac{1}{2}$.
\end{property}
{\it Proof.}
If $s_{i^*}\in S_0^*$, then
\[\max _ju_{i^*j}=\underline{V}(U)\leq \max _ju_{ij}.\]
On the other hand, from Property \ref{Pro1}, we have
\[ \max _ju_{ij}=\max _j(1-u_{ji})=1-\min_ju_{ji}\]
for all $i \in \{1,2,...,N\}$.
Therefore,
\[\overline{V}(U)=\min_ju_{ji^*} \geq \min_ju_{ji}.\]
That is to say $s_{i^*} \in S_1^*$ and $\underline{V}(U)=1-\overline{V}(U)$. Similarly, if $s_{i^*}\in S_1^*$, then $s_{i^*} \in S_0^*$. Therefore, $S_0^*=S_1^*=S^*$. If $\overline{V}(U)=\frac{1}{2}$, then $\underline{V}(U)=\frac{1}{2}$. From Lemma \ref{lemma2}, we can make a conclusion that $(s_{i},s_{j})$ is a Nash equilibium point if and only if $(s_{i},s_{j})\in S^*\times S^*$ and the Nash equilibrium outcome is $\frac{1}{2}$. $\blacksquare$

\begin{remark}
From Property \ref{pro3}, if $(s_{i_1},s_{j_1})$ and $(s_{i_2},s_{j_2})$ are Nash equilibrium points, then $(s_i,s_j)$ is a Nash equilibrium for all $i,j\in\{i_1,i_2,j_1,j_2\}$. Hence, in the case of multiple Nash equilibrium points, each player does not have to know which security strategy his opponent will use in the game, since all such strategies are in equilibrium and they yield the same value. Therefore, it is not necessary to solve all Nash equilibrium points. As a result, the aim of each player is to seek a strategy $s_{i^*}$ satisfied $u_{i^*j}\leq \frac{1}{2}$ for all $j\in V$.
\end{remark}

\begin{property}\label{pro4}
If the two players $l_0$ and $l_1$ pick the same strategy, then all followers converge to $\frac{y_0 +y_1}{2}$.
\end{property}
{\it Proof.} For the strategies pair $(s_i,s_i)$, it follows from (\ref{e3}) that
\[\alpha=\beta=(L+\diag\{s_i+s_i\})^{-1}s_i.\]
Because of $\alpha+\beta=\mathbf{1}_n$, we have $\alpha=\beta=\frac{1}{2}\mathbf{1}_n$.
Hence, from Lemma \ref{lemma3}, we can obtain that all followers converge to $\frac{y_0 +y_1}{2}$. $\blacksquare$
\begin{remark}
From Property \ref{pro3} and \ref{pro4}, it is obviously that if $\mathcal{G}^k(G_F)$ has a unique Nash equilibrium point, then $l_0$ and $l_1$ have a unique security strategy $s_{i^*}$. Consequently, all followers will achieve consensus.
\end{remark}

\subsection{Special case: $k=1$}
In this subsection we discuss the case of $k=1$ which means each player can connect only with one follower in $G_F$. Denote the game as $\mathcal{G}^1(G_F)$. Then the two players have $n$ alternatives and the strategies set can be denoted as $S=\{\mathbf{e}_1,\mathbf{e}_2,...,\mathbf{e}_n\}$. In order to find the optimal topology, we know from above Properties that it is necessary to determine whether the inequalities $u_{ij}\leq \frac{1}{2}$ holds or not, instead of to compute the precise value of $u_{ij}$. Then we have the following result:
\begin{theorem}\label{th1}
For the game $\mathcal{G}^1(G_F)$, we have $u_{ij}<(=~or~>)~\frac{1}{2} $ if and only if
\begin{equation}\label{uij1}
\parallel(L+diag\{\mathbf{e}_i\})^{-1}\mathbf{e}_j\parallel_1
<(= ~or~>)~\parallel(L+diag\{\mathbf{e}_j\})^{-1}\mathbf{e}_i\parallel_1.
\end{equation}
\end{theorem}
{\it Proof.}
From Property \ref{Pro0}, we have $u_{ij}<\frac{1}{2} \Leftrightarrow U_0(s_i,s_j)<U_1(s_i,s_j)$. According to (\ref{e3}) and (\ref{e5}), we get
 \[U_0(s_i,s_j)<U_1(s_i,s_j) \Leftrightarrow \mathbf{1}_n^T(L+\diag\{\mathbf{e}_i+\mathbf{e}_j\})^{-1}\mathbf{e}_j < \mathbf{1}_n^T(L+\diag\{\mathbf{e}_i+\mathbf{e}_j\})^{-1}\mathbf{e}_i.\]
It follows from
 $(L+\diag\{\mathbf{e}_i+\mathbf{e}_j\})^{-1}=
 \dfrac{\adj(L+\diag\{\mathbf{e}_i+\mathbf{e}_j\})}{\det(L+\diag\{\mathbf{e}_i+\mathbf{e}_j\})}$ that
 \[ \begin{aligned}
\mathbf{1}_n^T(L+\diag\{\mathbf{e}_i+\mathbf{e}_j\})^{-1}\mathbf{e}_j &< \mathbf{1}_n^T(L+\diag\{\mathbf{e}_i+\mathbf{e}_j\})^{-1}\mathbf{e}_i\\ \Leftrightarrow\\
\mathbf{1}_n^T\adj(L+\diag\{\mathbf{e}_i+\mathbf{e}_j\})\mathbf{e}_j &<
 \mathbf{1}_n^T\adj(L+\diag\{\mathbf{e}_i+\mathbf{e}_j\})\mathbf{e}_i.
 \end{aligned}\]
 Noticing that
  \begin{equation}\label{adjL}
 \adj(L+\diag\{\mathbf{e}_i+\mathbf{e}_j\})\mathbf{e}_j=\adj(L+\diag\{\mathbf{e}_i\})\mathbf{e}_j
\end{equation}
 and
\begin{equation}\label{adjL2}
\adj(L+\diag\{\mathbf{e}_i+\mathbf{e}_j\})\mathbf{e}_i=\adj(L+\diag\{\mathbf{e}_j\})\mathbf{e}_i,
\end{equation}
 one has
 \[
 \begin{aligned}
\mathbf{1}_n^T\adj(L+\diag\{\mathbf{e}_i+\mathbf{e}_j\})\mathbf{e}_j & <
 \mathbf{1}_n^T\adj(L+\diag\{\mathbf{e}_i+\mathbf{e}_j\})\mathbf{e}_i \\
 \Leftrightarrow\\
\mathbf{1}_n^T\adj(L+\diag\{\mathbf{e}_i\})\mathbf{e}_j &< \mathbf{1}_n^T\adj(L+\diag\{\mathbf{e}_j\})\mathbf{e}_i.
 \end{aligned}\]
By using Laplace expansion along column $k$, it follows from Lemma \ref{e-1} that
\begin{equation}\label{Lei}
\begin{aligned}
\det(L+\diag\{\mathbf{e}_i\})&=\sum_{k=1}^n(-1)^{i+k}l_{ki}\det L(k,i)+\det L(i,i)\\
&=\det L+\mathbf{e}_i^T\adj L ~\mathbf{e}_i=\tau(G_F)
\end{aligned}
\end{equation}
where $l_{ki}$ is the $ki^{th}$ entry of $L$.
 Likewise, we also have $\det(L+\diag\{\mathbf{e}_j\})=\tau(G_F)$. Thus, we obtain
\[\begin{aligned}
\mathbf{1}_n^T\adj(L+\diag\{\mathbf{e}_i\})\mathbf{e}_j &< \mathbf{1}_n^T\adj(L+\diag\{\mathbf{e}_j\})\mathbf{e}_i\\ \Leftrightarrow  \\
\mathbf{1}_n^T(L+\diag\{\mathbf{e}_i\})^{-1}\mathbf{e}_j &< \mathbf{1}_n^T(L+\diag\{\mathbf{e}_j\})^{-1}\mathbf{e}_i.
 \end{aligned}\]
Denote $\overline{L}_i=\left(
                   \begin{array}{cc}
                     1 & -\mathbf{e}_i^T \\
                     -\mathbf{e}_i & L+\diag\{\mathbf{e}_i\} \\
                   \end{array}
                 \right)
$. Due to $G_F$ is connected, $\overline{L}_i$ is also the Laplacian matrix of a connected graph by adding a vertex $0$ and an edge $(0,i)$ to $G_F$. Therefore, it follows from Lemma \ref{lemma L} that $(L+\diag\{\mathbf{e}_i\})^{-1}$ is a nonnegative matrix. Likewise, $(L+\diag\{\mathbf{e}_j\})^{-1}$ is a nonnegative matrix too. Then, we get
\[\mathbf{1}_n^T(L+\diag\{\mathbf{e}_i\})^{-1}\mathbf{e}_j
=\parallel(L+diag\{\mathbf{e}_i\})^{-1}\mathbf{e}_j\parallel_1\]
and
\[\mathbf{1}_n^T(L+\diag\{\mathbf{e}_j\})^{-1}\mathbf{e}_i
=\parallel(L+diag\{\mathbf{e}_j\})^{-1}\mathbf{e}_i\parallel_1.\]
Thus, we can make a conclusion that
\[
u_{ij}<\frac{1}{2} \Leftrightarrow \parallel(L+diag\{\mathbf{e}_i\})^{-1}\mathbf{e}_j\parallel_1
<\parallel(L+diag\{\mathbf{e}_j\})^{-1}\mathbf{e}_i\parallel_1.\]
Similarly, we can proof the case of "$=$" and "$>$". $\blacksquare$

Based on this result, we can decide whether an interaction graph $\tilde{G}(\mathbf{e}_i,\mathbf{e}_j)$ is the optimal topology or not. The following result is obtained:

\begin{theorem}
For the game $\mathcal{G}^1(G_F)$, let
\[
S_e=\{\mathbf{e}_{i^*}|~\mathbf{e}_{i^*}\in S,~\parallel(L+diag\{\mathbf{e}_{i^*}\})^{-1}\mathbf{e}_k\parallel_1
\leq\parallel(L+diag\{\mathbf{e}_k\})^{-1}\mathbf{e}_{i^*}\parallel_1,~~k\in \mathcal{I}_n\}.
\] 
Then $\tilde{G}(\mathbf{e}_{i^*},\mathbf{e}_{j^*})$ is the optimal topology if and only if $\mathbf{e}_{i^*} \in S_e$ and $\mathbf{e}_{j^*} \in S_e$.
\end{theorem}
{\it Proof.} Because of Definition \ref{optimal}, it is suffice to prove that $(\mathbf{e}_{i^*},\mathbf{e}_{j^*})$ is a Nash equilibium point if and only if $\mathbf{e}_{i^*} \in S_e$ and $\mathbf{e}_{j^*} \in S_e$. It follows from Theorem \ref{th1} that
\begin{equation}\label{e11}
\mathbf{e}_{i^*} \in S_e \Leftrightarrow u_{i^*k}\leq \frac{1}{2},~~ k\in \mathcal{I}_n.
\end{equation}
Sufficiency. Due to (\ref{e11}), we have $\max_k u_{i^*k}=\max_k u_{j^*k}=\frac{1}{2}$. Then, it follows from Property \ref{Pro1} that $\overline{V}(U)=\frac{1}{2}$ and moreover, $\mathbf{e}_{i^*}$ and $\mathbf{e}_{j^*}$ are security strategies. Hence, from Property \ref{pro3}, we have $(\mathbf{e}_{i^*},\mathbf{e}_{j^*})$ is a Nash equilibrium point.

Necessity. Firstly, we will prove $u_{i^*j^*}=\frac{1}{2}$ by contradiction. Since $(\mathbf{e}_{i^*},\mathbf{e}_{j^*})$ is a Nash equilibrium point, one has $u_{i^*j}\leq u_{i^*j^*} \leq u_{ij^*}$ for all $i,j\in \mathcal{I}_n$. Suppose $u_{i^*j^*}>\frac{1}{2}$. Recalling $u_{j^*j^*}=\frac{1}{2}$, we get $u_{i^*j^*}>u_{j^*j^*}$ which conflicts with the fact that $u_{i^*j^*} \leq u_{ij^*}$. Supposing $u_{i^*j^*}<\frac{1}{2}$, we have $u_{i^*i^*} > u_{ij^*}$ which has a conflict with $u_{i^*j}\leq u_{i^*j^*}$. Therefore, we have proved $u_{i^*j^*}=\frac{1}{2}$. Then it follows that $u_{i^*j}\leq \frac{1}{2}\leq u_{ij^*}$ for all $i,j\in \mathcal{I}_n$. According (\ref{e11}), we have $\mathbf{e}_{i^*} \in S_e$. Since $u_{ij^*}=1-u_{j^*i}$, we obtain $u_{j^*i}\leq \frac{1}{2}$. Hence, $\mathbf{e}_{j^*} \in S_e$.
$\blacksquare$

For the game $\mathcal{G}^1(G_F)$, we have an interesting property as follows:
\begin{property}\label{Pro5}
For every strategies pair $(\mathbf{e}_i,\mathbf{e}_j)\in S \times S$ of $\mathcal{G}^1(G_F)$, we always have
\[\lim_{t\rightarrow\infty}(x_i(t)-y_0)=\lim_{t\rightarrow\infty}(y_1-x_j(t))\]
\end{property}
{\it Proof.} For the case of $i=j$, it follows from Property \ref{pro4} that $\lim_{t\rightarrow\infty}(x_i(t)-y_0)=\lim_{t\rightarrow\infty}(y_1-x_j(t))=\frac{y_1-y_0}{2}$.

For the case of $i\neq j$, one can renumber nodes of $G_F$ by exchanging the number of $1$ and $i$ and that of $2$ and $j$, i.e., $1 \leftrightarrow i, ~ 2\leftrightarrow j$. Then, it follows that the the new Laplacian matrix $L'=PLP^T$ where
 \[P=[\mathbf{e}_i,\mathbf{e}_j,...\mathbf{e}_{i-1},\mathbf{e}_1,\mathbf{e}_{i+1},
 ...,\mathbf{e}_{j-1},\mathbf{e}_2,\mathbf{e}_{j+1},...,\mathbf{e}_n]\]
 is orthogonal.
 It is easy to prove that
 \[
 \mathbf{1}_n^T\adj(L+\diag\{\mathbf{e}_i\})\mathbf{e}_j
 =\mathbf{1}_n^T\adj(L'+\diag\{\mathbf{e}_1\})\mathbf{e}_2,\]
 and
 \[\mathbf{1}_n^T\adj(L+\diag\{\mathbf{e}_j\})\mathbf{e}_i
 =\mathbf{1}_n^T\adj(L'+\diag\{\mathbf{e}_2\})\mathbf{e}_1.\]
 Consequently, it suffices to prove the theorem in the case of $i=1$ and $j=2$.

Let $\tilde{L}_1=L+diag\{\mathbf{e}_1\}$ and $\tilde{L}_2=L+diag\{\mathbf{e}_2\}$, it follows that
\[\tilde{L}_1=\tilde{L}_2+diag\{1,-1,0,...,0\}.\]
Denote $\tilde{L}_1^{-1}=(\mathbf{p}_1,\mathbf{p}_2,...,\mathbf{p}_n)$ and $\tilde{L}_2^{-1}=(\mathbf{r}_1,\mathbf{r}_2,...,\mathbf{r}_n)$ where $\mathbf{p}_i,~\mathbf{r}_i$ are n-dimension vectors. Thus, it is easy to find that
\[\begin{aligned}
&\tilde{L}_1\tilde{L}_2^{-1}=I_n+(\mathbf{r}_1,-\mathbf{r}_2,\mathbf{0}_n,...,\mathbf{0}_n)^T,\\
&\tilde{L}_2\tilde{L}_1^{-1}=(\tilde{L}_1\tilde{L}_2^{-1})^{-1}=I_n+(-\mathbf{p}_1,\mathbf{p}_2,\mathbf{0}_n,...,\mathbf{0}_n)^T,
\end{aligned}\]
and $\mathbf{p}_1=\mathbf{r}_2=\mathbf{1}_n$. Consequently, $\tilde{L}_1\tilde{L}_2^{-1}$ and $\tilde{L}_2\tilde{L}_1^{-1}$ have eigenvalues $\lambda_1, \lambda_2, 1,...,1$ and $\lambda_1^{-1}, \lambda_2^{-1}, 1,...,1$, respectively. Noticing from (\ref{Lei}) that
\[\lambda_1\lambda_2=\mid\tilde{L}_1\tilde{L}_2^{-1}\mid
=\mid\tilde{L}_1\mid\mid\tilde{L}_2\mid^{-1}=1,\]
we have $\lambda_2=\lambda_1^{-1}$. Then the eigenvalues of $\tilde{L}_2\tilde{L}_1^{-1}$  are same as those of $\tilde{L}_1\tilde{L}_2^{-1}$. Letting $\mathbf{p}_2=(w_1,w_2,...,w_n)^T$ and $\mathbf{r}_1=(z_1,z_2,..,z_n)$, we obtain
\[\left|\lambda I_n-\tilde{L}_2\tilde{L}_1^{-1}\right|
=(\lambda-1)^{n-2}
\left| \begin{array}{cc}
 \lambda & 1 \\[-1.5em]
 -w_1~~~ & \lambda-w_2-1 \\
  \end{array}\right|
=(\lambda-1)^{n-2}(\lambda^2-(w_2+1)\lambda+w_1)
 \]
and
\[\left|\lambda I_n-\tilde{L}_1\tilde{L}_2^{-1}\right|
=(\lambda-1)^{n-2}
\left| \begin{array}{cc}
\lambda-z_1-1  &~~~ -z_2 \\[-1.5em]
1 & \lambda \\
 \end{array}\right|
=(\lambda-1)^{n-2}(\lambda^2-(z_1+1)\lambda+z_2).
 \]
Thus one has $w_1=z_2=1$ and $w_2=z_1=\lambda_1 +\lambda_1^{-1}$.
From Lemma \ref{lemma3}, recalling (\ref{adjL}), (\ref{adjL2}) and (\ref{Lei}) we get
 \[\begin{aligned}
\lim_{t\rightarrow\infty}X_f(t)
 &=(L+\diag\{\mathbf{e}_1+\mathbf{e}_2\})^{-1}(\mathbf{e}_1,\mathbf{e}_2)(y_0,y_1)^T\\
&=\dfrac{\tau(G_F)}{|L+\diag\{\mathbf{e}_1+\mathbf{e}_2\}|}
\left[\tilde{L}_2^{-1}\mathbf{e}_1,
\tilde{L}_1^{-1}\mathbf{e}_2\right](y_0,y_1)^T\\
&=\dfrac{\tau(G_F)}{|L+\diag\{\mathbf{e}_1+\mathbf{e}_2\}|}[\mathbf{r}_1,\mathbf{p}_2](y_0,y_1)^T.
\end{aligned}\]
Denote $\mu=\dfrac{\tau(G_F)}{|L+\diag\{\mathbf{e}_1+\mathbf{e}_2\}|}$. It follows that
\[
\lim_{t\rightarrow\infty}x_1(t)=\mu y_0+\mu(\lambda_1 +\lambda_1^{-1})y_1~~ \text{and}~~
\lim_{t\rightarrow\infty}x_2(t)=\mu(\lambda_1 +\lambda_1^{-1})y_0+\mu y_1.
\]
Since $\mu+\mu(\lambda_1 +\lambda_1^{-1})=1$, it is easy to prove
\[\lim_{t\rightarrow\infty}(x_i(t)-y_0)=\lim_{t\rightarrow\infty}(y_1-x_j(t)). ~~~\blacksquare \]
\subsection{Graphical results}
In this subsection we will deduce some graphical results for $\mathcal{G}^1(G_F)$. Firstly, we present the following Lemma.
\begin{lemma}
For all $i \neq j$, we have
\begin{equation}\label{e12}
\mathbf{1}_n^T\adj(L+\diag\{\mathbf{e}_i\})\mathbf{e}_j=n\tau(G_F)+ M^{(ij)}
\end{equation}
where
\[M^{(ij)}=\left|\begin{array}{cccccccc}
             l_{11} & \cdots & l_{1,i-1}& l_{1,i+1} &  \cdots& l_{1j}& \cdots & l_{1n} \\[-1.5em]
              \cdots &  \cdots &  \cdots &  \cdots &  \cdots &  \cdots &  \cdots &  \cdots \\[-1.5em]
             l_{i-1,1} & \cdots & l_{i-1,i-1} & l_{i-1,i+1} & \cdots &l_{i-1,j}&  \cdots & l_{i-1,n}\\[-1.5em]
              l_{i+1,1} & \cdots & l_{i+1,i-1} & l_{i+1,i+1} & \cdots & l_{i+1,j}&  \cdots & l_{i+1,n} \\[-1.5em]
             \cdots &  \cdots &  \cdots &  \cdots &  \cdots &  \cdots &  \cdots &  \cdots \\[-1.5em]
             l_{j-1,1} & \cdots & l_{j-1,i-1} &l_{j-1,i+1} &\cdots &l_{j-1,j}& \cdots  & l_{j-1,n} \\[-1.5em]
             1 &  \cdots & 1 &  1& \cdots & 1 &  \cdots & 1 \\[-1.5em]
             l_{j+1,1} & \cdots & l_{j+1,i-1} &l_{j+1,i+1} &\cdots &l_{j+1,j}& \cdots &l_{j+1,n} \\[-1.5em]
             \cdots &  \cdots &  \cdots &  \cdots &  \cdots &  \cdots &  \cdots &  \cdots \\[-1.5em]
             l_{n1} & \cdots & l_{n,i-1} &l_{n,i+1} &\cdots & l_{nj}&  \cdots & l_{nn}\\
           \end{array}\right|.\]

\end{lemma}
{\it Proof.} Firstly, we will prove the case of $i<j$. Denote $\tilde{L}=L+diag\{\mathbf{e}_i\}$ and $\Delta_i=diag\{\mathbf{e}_i\}$. Then we have
\begin{equation}
\det \tilde{L}(j,k)= \det\left(L(j,k)+\Delta_i(j,k)\right)=\det L(j,k)+ M_{jk}
\end{equation}
where $M_{ji}=0$ and
\[
M_{jk}=\left|
           \begin{array}{cccccccc}
             l_{11} & \cdots & l_{1i}& \ldots &  l_{1,k-1} & l_{1,k+1}& \cdots & l_{1n} \\[-1.5em]
              \cdots &  \cdots &  \cdots &  \cdots &  \cdots &  \cdots &  \cdots &  \cdots \\[-1.5em]
               l_{i-1,1} & \cdots & l_{i-1,i} & \cdots & l_{i-1,k-1} & l_{i-1,k+1}&  \cdots & l_{i-1,n} \\[-1.5em]
             0 &  \cdots & 1 &  \cdots & 0 & 0 &  \cdots & 0 \\[-1.5em]
              l_{i+1,1} & \cdots & l_{i+1,i} & \cdots & l_{i+1,k-1} & l_{i+1,k+1}&  \cdots & l_{i+1,n} \\[-1.5em]
             \cdots &  \cdots &  \cdots &  \cdots &  \cdots &  \cdots &  \cdots &  \cdots \\[-1.5em]
             l_{j-1,1} & \cdots & l_{j-1,i} & \cdots & l_{j-1,k-1} & l_{j-1,k+1}&  \cdots & l_{j-1,n} \\[-1.5em]
             l_{j+1,1} & \cdots & l_{j+1,i} & \cdots & l_{j+1,k-1} & l_{j+1,k+1}&  \cdots & l_{j+1,n} \\[-1.5em]
             \cdots &  \cdots &  \cdots &  \cdots &  \cdots &  \cdots &  \cdots &  \cdots \\[-1.5em]
             l_{n1} & \cdots & l_{ni} & \cdots & l_{n,k-1} & l_{n,k+1}&  \cdots & l_{nn} \\
           \end{array}
         \right|
\]
for all $k\neq i$. It follows that
\[
\mathbf{1}_n^T\adj(L+\diag\{\mathbf{e}_i\})\mathbf{e}_j=\sum_{k=1}^n(-1)^{k+j}\det \tilde{L}(j,k)=n\tau(G_F)+ \sum_{k=1}^n(-1)^{k+j} M_{jk}.
\]
It is easy to prove that
\[\sum_{k=1}^n(-1)^{k+j} M_{jk}=\sum_{k=1}^n\left|
           \begin{array}{ccccccccc}
             l_{11} & \cdots & l_{1,i-1}&l_{1,i+1}& \ldots &  l_{1,k} & l_{1,k+1}& \cdots & l_{1n} \\[-1.5em]
              \cdots &  \cdots &  \cdots&  \cdots &  \cdots &  \cdots &  \cdots &  \cdots &  \cdots \\[-1.5em]
               l_{i-1,1} & \cdots & l_{i-1,i-1}& l_{i-1,i+1} & \cdots & l_{i-1,k} & l_{i-1,k+1}&  \cdots & l_{i-1,n} \\[-1.5em]
               l_{i+1,1} & \cdots & l_{i+1,i-1} & l_{i+1,i+1}& \cdots & l_{i+1,k} & l_{i+1,k+1}&  \cdots & l_{i+1,n} \\[-1.5em]
             \cdots &  \cdots &  \cdots &  \cdots & \cdots &  \cdots &  \cdots &  \cdots &  \cdots \\[-1.5em]
             l_{j-1,1} & \cdots & l_{j-1,i-1}& l_{j-1,i+1} & \cdots & l_{j-1,k} & l_{j-1,k+1}&  \cdots & l_{j-1,n} \\[-1.5em]
             0 &  \cdots &  0&  0 &  \cdots &  1 &  0 &  \cdots &  0 \\[-1.5em]
             l_{j+1,1} & \cdots & l_{j+1,i-1} & l_{j+1,i+1}& \cdots & l_{j+1,k} & l_{j+1,k+1}&  \cdots & l_{j+1,n} \\[-1.5em]
             \cdots &  \cdots &  \cdots &  \cdots & \cdots &  \cdots &  \cdots &  \cdots &  \cdots \\[-1.5em]
             l_{n1} & \cdots & l_{n,i-1}& l_{n,i+1} & \cdots & l_{n,k} & l_{n,k+1}&  \cdots & l_{nn} \\
           \end{array}
         \right|=M^{(ij)}
\]
Hence, we obtain
 \[\mathbf{1}_n^T\adj(L+\diag\{\mathbf{e}_i\})\mathbf{e}_j=n\tau(G_F)+ M^{(ij)}.\]
Similarly, we can prove the case of $i>j$. $\blacksquare$

Now we will give some graphical results for the game $\mathcal{G}^1(G_F)$:
\begin{theorem}\label{special nodes}
For two nodes $i,j\in V$, if $\mathcal{N}_i\backslash \{j\}\supseteq \mathcal{N}_j\setminus\{i\}$, then $u_{ij}\leq\frac{1}{2}$ and the equality holds if and only if $\mathcal{N}_i\setminus\{j\}=\mathcal{N}_j\setminus\{i\}$.
\end{theorem}
{\it Proof.} Similar to the proof of Property \ref{Pro5}, it suffices to prove the theorem in the case of $i=1$ and $j=2$.

It follows from (\ref{uij1}) and (\ref{e12}) that $u_{12}\leq \frac{1}{2}$ if and only if $ \det M^{(21)}-\det M^{(12)}\geq 0$. Since (\ref{e12}), we have
     \[\det M^{(21)}-\det M^{(12)}=\left|
                   \begin{array}{cccc}
                     0 & 1 & \cdots & 1 \\[-1.5em]
                     l_{31}-l_{32} & l_{33} & \cdots  & l_{3n} \\[-1.5em]
                     \cdots & \cdots  & \cdots  & \cdots  \\[-1.5em]
                     l_{31}-l_{n2}& l_{n3} & \cdots & l_{nn} \\
                   \end{array}
                 \right|.\]

 If $\mathcal{N}_1\backslash \{2\}= \mathcal{N}_2\setminus\{1\}$, then $l_{j1}-l_{j2}=0$, $j\in\{3,4,...,n\}$. It follows that $\det M^{(21)}-\det M^{(12)}=0$. Hence $u_{12}=\frac{1}{2}.$

 If $\mathcal{N}_1\backslash \{2\} \supset \mathcal{N}_2\setminus\{1\}$, denote $\mathcal{N}_2\backslash \{1\}=\{i_1,i_2,...,i_k\}$ and $\mathcal{N}_1\backslash \{2\}=\{i_1,i_2,...,i_k,i_{k+1},...,i_{k+h}\}$, then $l_{j1}-l_{j2}=0$ for $j~\notin~\{1,2,i_{k+1},...,i_{k+h}\}$ and $l_{j1}-l_{j2}=-1$ for $j~\in~\{i_{k+1},...,i_{k+h}\}$. Therefore,
     \[\det M^{(21)}-\det M^{(12)}=\sum_{m=k+1}^{k+h}\sum_{j=3}^n(-1)^{j+i_m}Q_{i_mj}, \]
where
\[Q_{i_mj}=
      \left|
        \begin{array}{cccccc}

          l_{33} & \cdots & l_{3,j-1} & l_{3,j+1}&\cdots & l_{3n} \\[-1.5em]
          \cdots &\cdots & \cdots & \cdots & \cdots & \cdots \\[-1.5em]
          l_{i_m-1,3}&\cdots & l_{i_m-1,j-1} & l_{i_m-1,j+1} & \cdots & l_{i_m-1,n} \\[-1.5em]
          l_{i_m+1,3}&\cdots & l_{i_m+1,j-1} & l_{i_m+1,j+1}& \cdots & l_{i_m+1,n} \\[-1.5em]
          \cdots &\cdots & \cdots & \cdots & \cdots & \cdots \\[-1.5em]
          l_{n3} &\cdots & l_{n,j-1} & l_{n,j+1}& \cdots & l_{nn}  \\
          \end{array}
      \right|.
\]
 It is easy to observe that $(-1)^{j+i_m}Q_{i_mj}$ is the $(j-2,i_m-2)$ entry of the adjugate matrix of $L_{[3,4,...,n]}$. From Lemma \ref{lemma L}, we know that $L_{[3,4,...,n]}$ is positive definite and $L_{[3,4,...,n]}^{-1}$ is a nonnegative matrix. Then it follows that the adjugate matrix of $L_{[3,4,...,n]}$ is an invertible nonnegative matrix. Hence, for all $j\in \{3,4,...,n\}$ and $m \in \{k+1,k+2,...,k+h\}$, we have $(-1)^{j+i_m}Q_{i_mj}\geq 0$ and the inequality holds at least with one $j\in \{3,4,...,n\}$ which implies that $\det M^{(21)}-\det M^{(12)}>0.$ Consequently, we can make a conclusion that $u_{12}<\frac{1}{2}.$ $\blacksquare$

\begin{corollary}
If $\mathcal{N}_i=\{j\}$, then $u_{ij}\geq\frac{1}{2}$ and the equality holds if and only if $n=2$.
\end{corollary}
{\it Proof.} Since $\mathcal{N}_i\setminus \{j\}=\emptyset$, the proof is straightforward. $\blacksquare$

\begin{theorem} \label{special graph}
For the game $\mathcal{G}^1(G_F)$,
\begin{enumerate}
  \item if $G_F$ is a circulant graph, then the graph $\tilde{G}(\mathbf{e}_i,\mathbf{e}_j)$ is the optimal topology for every $(\mathbf{e}_i,\mathbf{e}_j)\in S\times S$;
  \item if $G_F$ has a vertex $i^c$ satisfied $\mathcal{N}_{i^c}=V\backslash \{i^c\}$, then $\tilde{G}(\mathbf{e}_{i^c},\mathbf{e}_{i^c})$ is the optimal topology.
\end{enumerate}
\end{theorem}
{\it Proof.}
Firstly, we will prove (1):

If $G_F$ is a circulant graph, then the adjacent matrix $A$ is a circulant matrix. It follows that $L$ is also a circulant matrix. Without loss of generality, we assume that $i<j$. Denote $\tilde{L}_i=L+diag\{\mathbf{e}_i\}$ and $\tilde{L}_j=L+diag\{\mathbf{e}_j\}$. Then for a permutation matrix
\[P=\left(
      \begin{array}{ccc}
      I_{i-1}    & \mathbf{0} & \mathbf{0} \\[-1.5em]
     \mathbf{0}  & \mathbf{0} & I_{n+1-j} \\[-1.5em]
      \mathbf{0} &  I_{j-i}   & \mathbf{0} \\
      \end{array}
    \right),
\]
we have $\tilde{L}_i=P\tilde{L}_jP^T$.
Notice that $P$ is orthogonal, we obtain that
\[\mathbf{1}_n^T(\tilde{L}_i)^{-1}\mathbf{e}_j
=\mathbf{1}_n^TP(\tilde{L}_j)^{-1}P^T\mathbf{e}_j.\]
Then it follows from $P\mathbf{e}_j=\mathbf{e}_i$ and $\mathbf{1}_n^TP=\mathbf{1}_n^T$ that
\[\mathbf{1}_n^T(\tilde{L}_i)^{-1}\mathbf{e}_j=\mathbf{1}_n^T(\tilde{L}_j)^{-1}\mathbf{e}_i.\]
Hence, form Theorem \ref{th1}, we get $u_{ij}=\frac{1}{2}$ for all $i,j\in V$, i.e., $U=\frac{1}{2} \mathbf{1}_n\mathbf{1}_n^T$. Then we can deduce that the strategy set is $S^*=S$. In another words, the graph $\tilde{G}(\mathbf{e}_i,\mathbf{e}_j)$ is the optimal topology for every $(\mathbf{e}_i,\mathbf{e}_j)\in S\times S$.

Next, we will give the proof of (2):

For simplicity, we may take $i^c=n$. Owing to $\mathcal{N}_{n}=\{1,2,...,n-1\}$, one has $\mathcal{N}_j\setminus \{n\}\subseteq \mathcal{N}_{n}\setminus \{j\}$ for all $ j\in \{1,2,...,n-1\}$. Then from Theorem \ref{special nodes}, we have $u_{nj}\leq \dfrac{1}{2}$for all $ j\in \{1,2,...,n-1\}$. Hence, $\overline{V}(U)=\underline{V}(U)=\dfrac{1}{2}$ and $\mathbf{e}_n\in S^*$. From Property \ref{pro3} and \ref{pro4}, it follows that the strategies pair $(\mathbf{e}_n,\mathbf{e}_n)$ is a Nash equilibrium point. Thus, we can conclude that $\tilde{G}(\mathbf{e}_n,\mathbf{e}_n)$ is the optimal topology.  $\blacksquare$

\begin{remark}
The results of this theorem can give some theoretic explanations for reality. Firstly, the follower's interaction graph is circulant implies that the influence power of every follower to the others is equal. Therefore, everyone is the optimal strategy to both two players. Secondly, if a member has influence on all the others in a social network, then his influence power is largest. As a result, both two players will connect with him to minimize their payoff.    
\end{remark}
From Theorem \ref{special graph}, we can deduce the following results:
\begin{corollary}
If $G_F$ is a star graph with the center vertex $i^c$, then the system has a unique optimal topology $\tilde{G}(\mathbf{e}_{i^c},\mathbf{e}_{i^c})$. Moreover, all followers can achieve consensus, and the consensus state is $\frac{y_1+y_0}{2}$.
\end{corollary}
\begin{corollary}
 If $G_F$ is a complete graph, then $\tilde{G}(\mathbf{e}_i,\mathbf{e}_j)$ is the optimal topology for arbitrary $(\mathbf{e}_i,\mathbf{e}_j)\in S \times S$.
\end{corollary}

\section{Simulation}\label{s4}
In this section, we give two numerical simulations to illustrate the effectiveness of theoretical results in Section \ref{s3}.

\begin{figure}[h]
  \centering
  \includegraphics[scale=0.4]{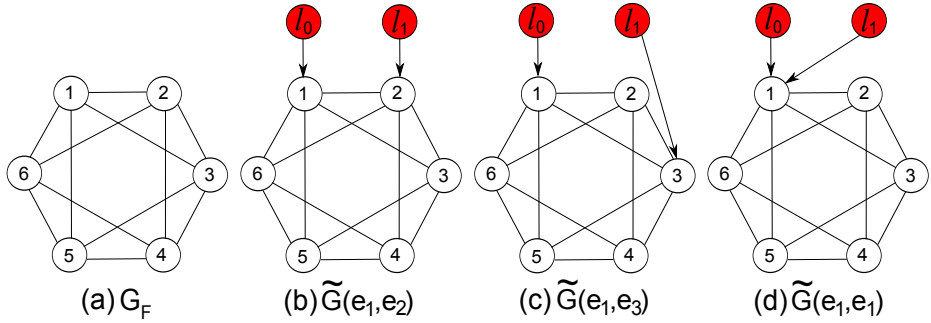}\\
  \caption{the intercation graph of Example 1}\label{fig3}
\end{figure}

\begin{figure}[hc]
  \centering
  \includegraphics[scale=0.5]{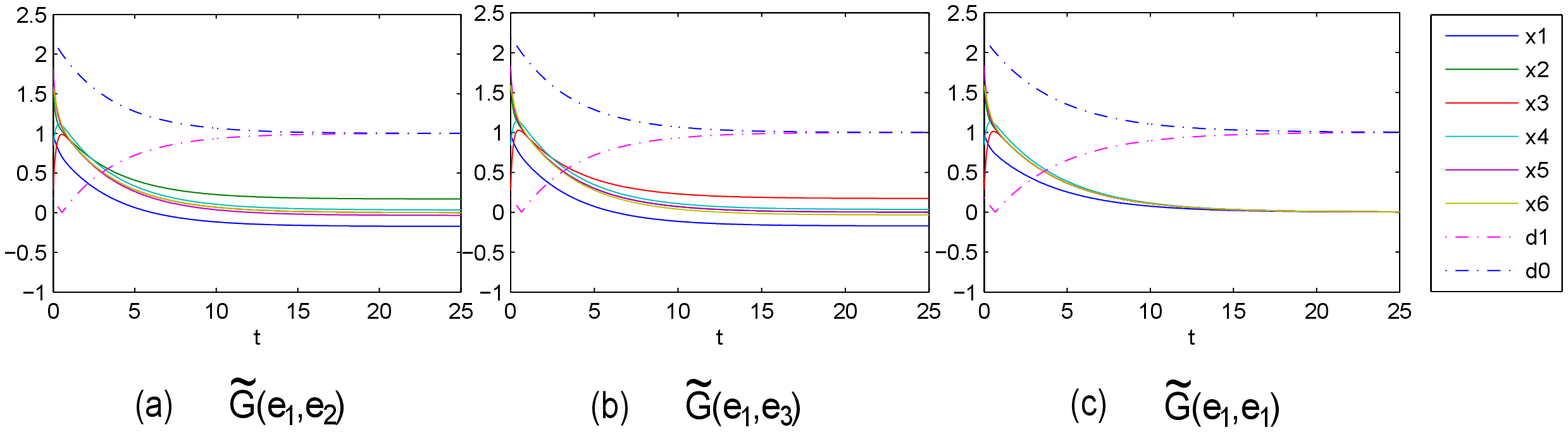}\\
  \caption{the followers' states and average distance under different strategies pairs}\label{fig4}
\end{figure}

\begin{figure}
  \centering
  \includegraphics[scale=0.4]{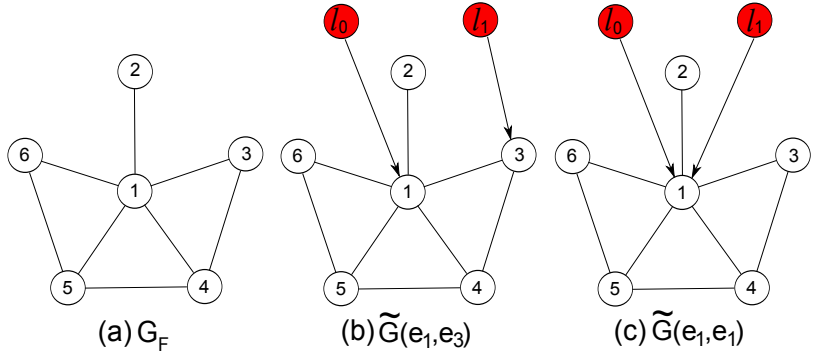}\\
  \caption{the intercation graph of Example 2}\label{fig5}
\end{figure}

\begin{figure}
  \centering
  \includegraphics[scale=0.4]{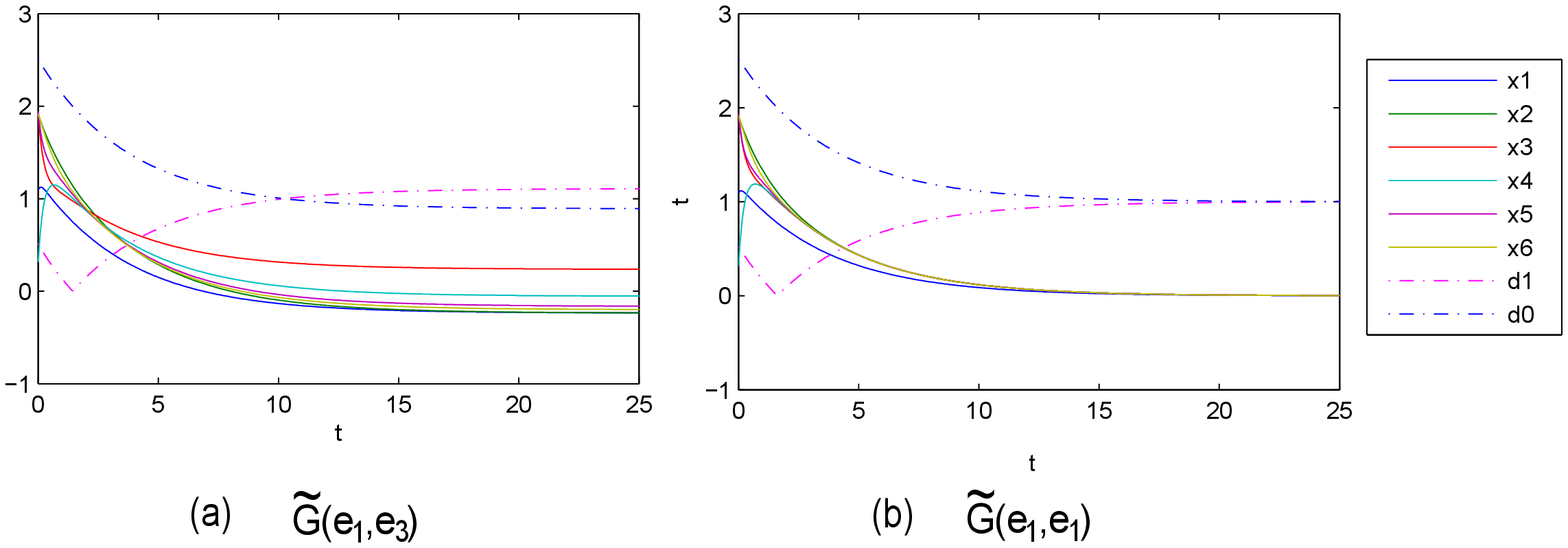}\\
  \caption{the followers' states and average distance under different strategies pairs}\label{fig6}
\end{figure}

Consider there are 6 followers labeled as 1-6 and two leaders labeled as $l_0$ and $l_1$. Let the leaders' initial state be -1 and 1, respectively. Define $d_0(t)= \frac{1}{6}\sum_{i=1}^6|x_i+1|$ and $d_1(t)= \frac{1}{6}\sum_{i=1}^6|1-x_i|$ as the average distance function of $l_0$ and $l_1$, respectively. Obviously, we have $\lim_{t\rightarrow\infty}d_j(t)=U_j$ ($j=0,1$).
\begin{example}
Let the followers' interaction graph $G_F$ be a circulant graph depicted in Fig. \ref{fig3} (a). Then, it is easy to obtain that the outcome matrix of the game $\mathcal{G}^1(G_F)$ is $U=\frac{1}{2} \mathbf{1}_6\mathbf{1}_6^T$. Therefore we can conclude that all 36 strategies pairs are Nash equilibrium points. Consider three strategies pairs $(\mathbf{e}_1,\mathbf{e}_2)$, $(\mathbf{e}_1,\mathbf{e}_3)$ and $(\mathbf{e}_1,\mathbf{e}_1)$. The corresponded interaction graphs $\tilde{G}(\mathbf{e}_1,\mathbf{e}_2)$, $\tilde{G}(\mathbf{e}_1,\mathbf{e}_3)$ and $\tilde{G}(\mathbf{e}_1,\mathbf{e}_1)$ are described in Fig. \ref{fig3} (b), Fig. \ref{fig3} (c) and Fig. \ref{fig3} (d), respectively. Fig. \ref{fig4} shows the followers' states, $d_0(t)$ and $d_1(t)$ under $\tilde{G}(\mathbf{e}_1,\mathbf{e}_2)$, $\tilde{G}(\mathbf{e}_1,\mathbf{e}_3)$ and $\tilde{G}(\mathbf{e}_1,\mathbf{e}_1)$, respectively. Note that under those three different graphs, both $d_0(t)$ and $d_1(t)$ which are depicted by dashed lines in Fig. \ref{fig4} converge to 1. This result illustrates the effectiveness of theoretical results in Theorem 4. For $\tilde{G}(\mathbf{e}_1,\mathbf{e}_1)$, we can find that all followers converge to the middle point of two leaders' initial states, which is consistent with the result of Property 4.

\end{example}
\begin{example}
The followers' interaction graph $G_F$ is depicted in Fig. \ref{fig5} (a). It is easy to find that the follower 1 is a center node in $G_F$. We can obtain the outcome matrix of the game $\mathcal{G}^1(G_F)$ as following:
\[U=\left(
     \begin{array}{cccccc}
        \textbf{0.5}~~~~~~&  \textbf{0.3889}~~  &  \textbf{0.4455}~~ &   \textbf{0.4712} ~~  & \textbf{0.4712}~~ &   \textbf{0.4455}~~\\[-1.5em]
 \textbf{ 0.6111}~~ &  0.5~~~~~~&  0.5526 ~~   &0.5753   & 0.5753   & 0.5526\\[-1.5em]
    \textbf{0.5545 } &  0.4474   & 0.5~~~~~~ &   0.5273  &  0.5246  &  0.5~~~~~~\\[-1.5em]
\textbf{0.5288} &   0.4247  &  0.4727 &   0.5~~~~~~ &  0.5~~~~~~ &  0.4754\\[-1.5em]
    \textbf{0.5288 } &  0.4247  &  0.4754  &  0.5~~~~~~ &   0.5~~~~~~&  0.4727\\[-1.5em]
     \textbf{0.5545 }  & 0.4474   & 0.5~~~~~~&   0.5246  &  0.5273  &  0.5~~~~~~\\
    \end{array}
   \right).
\]
Obviously, $u_{11}=\min_i \max _ju_{ij}=\min_j \max _iu_{ij}$. Therefore, we can conclude that $\tilde{G}(\mathbf{e}_1,\mathbf{e}_1)$ is the optimal topology, which is consistent with the result of Theorem 4. Two strategies pairs $(\mathbf{e}_1,\mathbf{e}_3)$ and $(\mathbf{e}_1,\mathbf{e}_1)$ and the corresponding graphs $\tilde{G}(\mathbf{e}_1,\mathbf{e}_3)$ and $\tilde{G}(\mathbf{e}_1,\mathbf{e}_1)$ are depicted in Fig. \ref{fig5}(a) and Fig. \ref{fig5}(b), respectively. Fig. \ref{fig6} shows the followers' states, $d_0(t)$ and $d_1(t)$ under $\tilde{G}(\mathbf{e}_1,\mathbf{e}_3)$ and $\tilde{G}(\mathbf{e}_1,\mathbf{e}_1)$, respectively. We can conclude from Fig \ref{fig6}(a) that $U_0<U_1$ for the strategies pair $(\mathbf{e}_1,\mathbf{e}_3)$, which illustrates the effectiveness of theoretical results in Theorem 3.
\end{example}

\section{Conclusion}

In this paper, we make use of game theory to tackle containment control problem with leaders. Firstly, we assumed that the followers' interaction subgraph $G_F$ was undirected and connected and every leader can independently select $k ~(\geq1)$ followers to connect with. By choosing his connected followers, each leader attempted to minimize his payoff which was defined as the average distance from himself to all followers. Then we proved that the sum of two payoffs is constant. Because every strategies pair corresponds to an interaction graph of the system, it was noted that to find a Nash equilibrium is equivalent to solve the optimal topology of the system. Intuitively, this is induced by the constant number of followers: if a leader has more followers then the other leader has fewer followers. The same applies to the distance between the followers and their leader. Secondly, we redefined this game as a standard two-player zero-sum game denoted as $\mathcal{G}^k(G_F)$ and obtained some properties for it. Thirdly, we further investigated the case of $k=1$. For the game $\mathcal{G}^1(G_F)$, the necessary and sufficient condition for an interaction graph to be the optimal topology was given. And if $G_F$ was a circulant graph or a graph with a center node, then the optimal topology was also obtained. This work puts containment control in a game theoretical framework, this perspective will foster the understanding of the interactions between leaders. Future work may consider this game for some MASs with constrains, such as MASs with switching topologies, MASs under measurement noises and MASs with quantized information transmission, etc.



\end{document}